\documentclass[journal]{IEEEtran}
\usepackage{amsmath,amsfonts}
\usepackage{algorithmic}
\usepackage{algorithm}
\usepackage{array}
\usepackage[caption=false,font=normalsize,labelfont=sf,textfont=sf]{subfig}
\usepackage{textcomp}
\usepackage{stfloats}
\usepackage{url}
\usepackage{verbatim}
\usepackage{graphicx}
\usepackage{cite}
\usepackage{tabularx}
\usepackage{booktabs}
\usepackage{multirow}
\usepackage{pgfplots}

\usepackage{xcolor, soul}
\newif\ifsubmission

\submissiontrue

\newcommand{\review}[1]{
    \ifsubmission
        {#1}
    \else
        \sethlcolor{yellow}
        \hl{#1}
    \fi
}

\begin{document}

\title{Classical and learned MR to pseudo-CT mappings for accurate transcranial ultrasound simulation}

\author{Maria~Miscouridou, José A. Pineda-Pardo, Charlotte J. Stagg, Bradley E. Treeby, and Antonio Stanziola
\thanks{M. Miscouridou, B. E. Treeby, and A. Stanziola are with the Department of Medical Physics and Biomedical Engineering, University College London, London, UK. J. A. Pineda-Pardo is with HM CINAC, Fundación HM Hospitales de Madrid, University Hospital HM Puerta del Sur. CEU-San Pablo University, Móstoles, Madrid, Spain. C. J. Stagg is with the Wellcome Centre for Integrative Neuroimaging, FMRIB, Nuffield Department of Clinical Neuroscience, University of Oxford, Oxford, UK.
For the purpose of open access, the author has applied a Creative Commons Attribution (CC BY) license to any Author Accepted Manuscript version arising.}}



\maketitle

\begin{abstract}
Model-based treatment planning for transcranial ultrasound therapy typically involves mapping the acoustic properties of the skull from an x-ray computed tomography (CT) image of the head. Here, three methods for generating pseudo-CT images from magnetic resonance (MR) images were compared as an alternative to CT. A convolutional neural network (U-Net) was trained on paired MR-CT images to generate pseudo-CT images from either T1-weighted or zero-echo time (ZTE) MR images (denoted tCT and zCT, respectively). A direct mapping from ZTE to pseudo-CT was also implemented (denoted cCT). When comparing the pseudo-CT and ground truth CT images for the test set, the mean absolute error was 133, 83, and 145 Hounsfield units (HU) across the whole head, and 398, 222, and 336 HU within the skull for the tCT, zCT, and cCT images, respectively. Ultrasound simulations were also performed using the generated pseudo-CT images and compared to simulations based on CT. An annular array transducer was used targeting the visual or motor cortex. The mean differences in the simulated focal pressure, focal position, and focal volume were 9.9\%, 1.5 mm, and 15.1\% for simulations based on the tCT images, 5.7\%, 0.6 mm, and 5.7\% for the zCT, and 6.7\%, 0.9 mm, and 12.1\% for the cCT. The improved results for images mapped from ZTE highlight the advantage of using imaging sequences which improve contrast of the skull bone. Overall, these results demonstrate that acoustic simulations based on MR images can give comparable accuracy to those based on CT.


\end{abstract}


\section{\label{sec:introduction}Introduction}
Transcranial ultrasound therapy is a class of non-invasive techniques that leverage ultrasound energy to modify the structure or function of the brain, for example, to ablate brain tissue \cite{elias2016randomized}, modulate brain activity \cite{legon2014transcranial}, or deliver therapeutic agents through the blood-brain barrier \cite{gasca2021blood}. The major challenge is the delivery of ultrasound through the intact skull bone, which can significantly aberrate and attenuate the transmitted ultrasound waves. To counter the effect of the skull, computer simulations are often used to predict the intracranial pressure field \cite{deffieux2010numerical}, or to adjust phase delays to ensure a coherent focus \cite{sun1998focusing}. Conventionally, these simulations are based on acoustic material property maps extracted from x-ray computed tomography (CT) images \cite{aubry2003experimental}. For clinical treatments in a hospital environment, obtaining pre-treatment CT images doesn't necessarily pose significant challenges. However, for transcranial ultrasound stimulation (TUS), which is being widely explored as a neuroscientific tool in addition to its potential clinical applications \cite{darmani2022non}, obtaining CT images for healthy volunteer studies can be more problematic. In the current work, the use of pseudo-CT (pCT) images for computer simulations of transcranial ultrasound propagation in a TUS setting is investigated. Three different methods of pCT generation are compared: (1) using a deep neural network based on T1-weighted (T1w) magnetic resonance (MR) images, (2) using a deep neural network based on zero-echo time (ZTE) MR images, and (3) directly mapping from ZTE MR images using classical image processing techniques following Wiesinger \cite{wiesinger2018zero}.

The image-to-image translation (I2IT) of MR to pCT images of the brain and skull has been widely explored in the imaging literature, particularly in the context of PET-MR, where the pCT images are used for PET attenuation correction \cite{burgos2014attenuation}, and radiotherapy planning \cite{boulanger2021deep}. In many cases, deep learning has been shown to out-perform classical techniques \cite{han2017mr}. A variety of models and techniques have been used, including generative neural networks (GANs) \cite{zhu2017unpaired, isola2017image}, supervised learning \cite{zhang2020cross}, contrastive learning \cite{park2020contrastive, jiangtao2021mri} and denoising diffusion probabilistic models \cite{sasaki2021unit}. Often, the network architecture accounts for features at multiple scales using a U-Net \cite{han2017mr, presotto2022evaluation}. The inputs to the neural network can be the full 3D image volumes \cite{fu20192d3d}, 2D slices along one or multiple planes \cite{spadea2019deep}, or small 3D patches \cite{roy2017synthesizing}.

The performance of the trained network is heavily influenced by the quality of the training dataset and choice of loss function. Ideally, the loss function should be directly related to the final performance of interest: in our case, the acoustic properties of the generated fields from the pCT (see Sec. \ref{subsec:ultrasound_simulations}). However, in the absence of an efficient differentiable acoustic simulator, the corresponding training regime is often inefficient and losses are therefore defined in image space. Simple voxel-based metrics, such as mean absolute error (MAE) and mean squared error (MSE), are often used when registered image pairs are available \cite{martinez2021franken}. 

Unpaired images can also be used for successfully training I2IT models, often by relying on some form of cycle-consistency loss \cite{jabbarpour2022unsupervised, liu2021ct, arabi2019novel} implemented using a discriminator neural network \cite{creswell2018generative}. While theoretically effective and often capable of training complex models, such as ones that disentangle geometrical features and imaging modality \cite{wang2021disentangled}, training generative neural networks against a discriminator can be challenging and often unstable \cite{wiatrak2019stabilizing}.

Several previous works have investigated the use of learned pCTs for transcranial ultrasound simulation. In the context of high-intensity focused ultrasound (HIFU) ablation, Su \textit{et al} \cite{su2020transcranial} used a 2D U-Net trained on transverse slices from a dataset of 41 paired dual-echo ultrashort echo time (UTE) MR images and segmented CT images (only the segmented skull bone was used for training). Within the skull, the pCT images generated from the test set had an MAE of 105 $\pm$ 21 HU, and showed good correlation with CT in terms of skull thickness and skull density ratio. Coupled acoustic-thermal simulations were performed for a 1024 multi-element array and a deep brain target. Differences in the simulated acoustic field were not reported, but differences in the predicted peak temperature were less than 2 $^\circ$C.

In a related study, Liu \textit{et al} \cite{liu2022synthetic} used a 3D conditional generative adversarial network (cGAN) trained on patches from a dataset of 86 paired T1w and segmented CT skull images. Within the skull, the pCT images generated from the test set had an MAE of 191 $\pm$ 22 HU, and similarly showed good correlation with CT in terms of skull thickness and skull density ratio. Acoustic simulations for a 1024 multi-element array and deep brain target showed an average 23 $\pm$ 6.5 \% difference in the simulated intracranial spatial-peak pressure when using pCT vs CT, and 0.35 $\pm$ 0.40 mm difference in the focal position.

In the context of TUS, Koh \textit{et al} \cite{koh2021acoustic} used a 3D cGAN trained on 3D patches from a dataset of 33 paired T1w and CT images. The generated pCT images had an MAE of 86 $\pm$ 9 HU within the head, and 280 $\pm$ 24 HU within the skull. Acoustic simulations were performed using a single-element bowl transducer driven at 200 kHz targeted at three brain regions \review{(motor cortex, visual cortex, dorsal anterior cingulate cortex)}. Across all targets for the test set, the mean difference in the simulated intracranial peak pressure was 3.11 $\pm$ 2.79 \% and the mean difference in the focal position was 0.86 $\pm$ 0.52 mm. However, aberrations to the ultrasound waves are known to be significantly reduced at low transmit frequencies \cite{hynynen1998demonstration,gimeno2019experimental}, and thus the performance of learned pCTs at higher frequencies more commonly used for TUS remains an open question. 

Several studies have also explored directly mapping the skull acoustic properties from T1w \cite{wintermark2014t1}, UTE \cite{miller2015ultrashort,johnson2017improved,guo2019feasibility} or ZTE \cite{caballero2019zero} images. Wintermark \textit{et al} \cite{wintermark2014t1} generated virtual CTs from 3 different MR sequences and used a Bayesian segmentation strategy 
using a skull mask (obtained by CT thresholding) as a prior. Linear regression was used to estimate skull thickness and density, with the mappings from T1w images performing best. Acoustic simulations using calculated phase correction values from virtual CT showed good agreement with those from real CT. Phantom HIFU experiments were also performed where the difference in maximum temperature was 1 $^\circ$C when using MR-based correction compared to CT-based correction.

In a similar study, Miller \textit{et al} \cite{miller2015ultrashort} investigated using UTE images instead of CT to apply aberration corrections for HIFU treatments. Three ex vivo skull phantoms were imaged by UTE and CT, and the UTE was segmented into a binary skull mask. Experimental transcranial sonications were performed on each skull using aberration corrections derived from MR, CT, and using no corrections. The measured temperature rises with aberration corrections were 45 \% higher than non-corrected sonications, while there was no significant difference between the results from MR and CT-calculated corrections. 
UTE has also been shown to produce images with bone contrast highly correlated to that in CT images \cite{johnson2017improved}.

Guo \textit{et al} \cite{guo2019feasibility} also investigated the use of UTE images using a series of linear mappings and thresholding operations to derive pCT images. A linear regression model comparing skull density from UTE and CT showed they were highly correlated.
Acoustic properties of the skull derived from the UTE and CT images had less than 5 \% error and were used to run acoustic and thermal simulations. The temperature rise was 1 $^\circ$C higher in CT-based simulations compared to UTE simulations and the focal location was usually within 1 pixel (1.33 mm).

Finally, Caballero \textit{et al} \cite{caballero2019zero} extracted the bone from ZTE and CT images with a series of thresholding and morphological operators, and used the bone maps to extract skull measures, such as skull thickness and skull density ratio. It was demonstrated with linear regression that the skull measures derived from the two modalities highly correlate to each other, and also correlate with treatment efficiency.


The recent studies outlined above clearly demonstrate the feasibility of using pCT images for treatment planning in transcranial ultrasound therapy. Here, the previous results are extended in two ways. First, the ability to generate pCT images from T1w and ZTE images is directly compared using a unique dataset with high-resolution multi-modality imaging data. While T1w images are widely available, they suffer from poor skull contrast, particularly compared to specialised bone imaging sequences which enable visualisation of the bony anatomy \cite{wiesinger2018zero}. Second, acoustic simulations are performed in the context of TUS using a commonly-used transducer geometry and a 500 kHz transmit frequency for both motor \cite{legon2018transcranial,zeng2022induction} and visual \cite{lee2016transcranial} targets.



\section{Methods}

\subsection{Multi-modality data set}

\begin{figure}[t]
    \centering
    \includegraphics[width=0.5\textwidth]{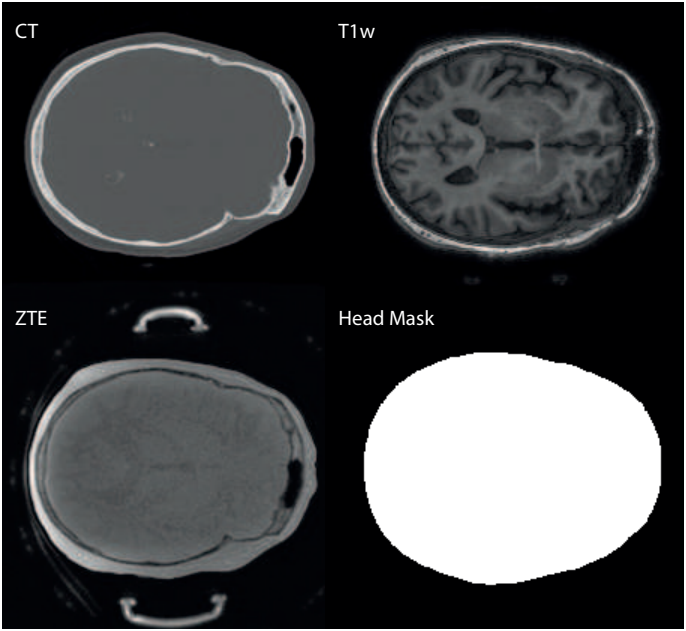}
    \caption{The input to the neural network is one or multiple adjacent transverse MR slices, acquired either using ZTE or T1w sequences. The output of the neural network is multiplied by a head mask and compared to the registered CT inside the mask.}
    \label{fig:inputs_examples}
\end{figure}

The dataset used for the study consisted of paired high-resolution CT and MR images. Subjects had previously been scheduled for transcranial MR-guided Focused Ultrasound Surgery (tcMRgFUS) thalamotomy. 
The study protocol was approved by the HM Hospitales Ethics Committee for Clinical Research and all participants provided written consent forms. Each subject had a CT and T1-weighted (fast-spoiled gradient echo) MR image, and a subset of them also had a ZTE MR image, giving a total of 171 paired CT-T1w datasets and 90 paired CT-ZTE datasets.
\review{The CT images were reconstructed using a bone-edge enhancement filter (FC30) and had a slice thickness of 1 mm and an in-plane resolution of approximately $0.45 \times 0.45$ mm. The in-plane resolution varied slightly between subjects due to the selected field-of-view (the number of pixels was fixed at $512 \times 512$).}

The MR images were acquired using a 3T GE Discovery 750 with an isotropic voxel size of 1 mm. Image acquisition parameters are described in detail in \cite{caballero2019zero}.

The images were processed as follows. First, bias-field correction was performed on the MR images using FAST \cite{zhang2001segmentation} to reduce image intensity non-uniformities resulting from transmit RF inhomogeneities and receive coil sensitivities. The CT images were then registered to the corresponding MR images using FLIRT \cite{jenkinson2001global, jenkinson2002improved}, using affine transformation with 12 degrees-of-freedom and a mutual-information cost function. As part of the registration step, CT images were resampled to match the resolution ($1$ mm isotropic) and field-of-view of the MR images. After registration, all volumes were padded to a cube with an edge length of 256 voxels. Following Han et al \cite{han2017mr}, the MR image intensities were then processed using midway histogram equalisation \cite{guillemot2016implementation}. The reference histogram was computed using images from the training set only (see Sec.\ \ref{sec:network_arch}), and was computed separately for the ZTE and T1w images. The images were also normalised using the mean and standard deviation of the corresponding training set.

To remove image artefacts (such as dental implants, equipment, headphones) and the bones outside the neurocranium (which are not relevant to TUS), a binary mask was manually annotated on the \verb=MNI152_T1_1mm= template brain \cite{fonov2009unbiased,fonov2011unbiased}. The MNI mask excluded the nasal and oral cavities, as well as everything below the foramen magnum. The MNI mask was then mapped to subject space by registering the MR images to the MNI template using FLIRT. The registered MNI mask was combined with a head mask (obtained by thresholding and filling the MR) to give a subject-specific mask that was used during training as described below. An example is given in Fig.\ \ref{fig:inputs_examples}. Note that the generation of the subject-specific mask doesn't require a ground truth CT, so it can be applied to any new subject data during inference.

\subsection{\label{sec:network_arch}Network architecture and training}

For mapping MR images to pCT, a 5-level U-Net \cite{unet2015} was implemented using Pytorch \cite{NEURIPS2019_9015} similar to the architecture reported by Han \cite{han2017mr}. Each level of the encoder consisted of either two (levels 1 and 2) or three (levels 3 to 5) convolutional layers with a rectified linear unit (ReLU) activation function followed by a batch normalisation layer. The convolutions used zero-padding, a $ 3 \times 3$ kernel, and a stride of 1. A max pooling layer was used between each level of the encoder, and skip connections between the encoder and decoder were used for the first four levels. The decoder was implemented as a mirrored version of the encoder with convolutional transpose (unpooling) layers \cite{zeiler2010deconvolutional} instead of max pooling layers. Both pooling and unpooling layers used a $2 \times 2$ kernel size and stride 2, followed by a dropout layer. The dropout probability $p_D$ was treated as a tunable hyperparameter which was optimised using the validation set. The best models for T1w and ZTE inputs were obtained with $p_D=0.1$ and $p_D=0$, respectively.

The input to the network was an $n \times 256 \times 256$ block of $n$ consecutive 2D transverse MR slices (see top-right and bottom-left of Fig.\  \ref{fig:inputs_examples} for an example of one slice). Using a small stack of images gives some 3D structural information to the network, as suggested in the discussion section of \cite{han2017mr}. The network output was always a single 2D transverse slice corresponding to the middle input slice, and 3D pCT volumes were reconstructed slice by slice. Input stack sizes of $n=1, 3, 5, 7, 9, 11$, and $15$ were used in preliminary testing. A stack size of $11$, i.e. $5$ additional image slices on either side of the primary input slice, gave the lowest MAE on the validation set. Notably, the use of multiple input slices significantly reduced the occurrence of skull discontinuities between slices in the 3D pCT images, particularly for the T1w inputs (this is discussed further in Sec.\ \ref{sec:results_pct}). A stack size of 11 was thus used in all subsequent training. 

The input was mapped to 64 channels in the first convolution. The number of channels was doubled at each layer of the encoder and halved in each layer of the decoder, except for the deep-most layer. A final $1 \times 1$ convolutional layer was implemented to map each 64-component feature vector from the previous layer to image voxels in Hounsfield units.

The network was trained separately for T1w and ZTE inputs. The ZTE dataset consisted of 90 subjects, with 62 subjects used for the training set $\Omega_T$, 14 for the validation set $\Omega_V$ and 14 for the testing set $\Omega_E$. Subjects were randomly assigned to each set. The T1w dataset consisted of 171 subjects. The same 14 subjects were used for the test set, with 26 subjects used for the validation set, and 131 for the training set. Data augmentation was also performed using random affine transformations with bilinear interpolation, with rotations in the range $\pm 10 ^\circ$, translations between $\pm$ 5 \% of the image size, and shears parallel to the x axis of $\pm 2.5 ^\circ$.

The network $f_\theta$ was trained by minimising the average $\ell_1$ norm of the masked error over the training set
\begin{equation}
    \mathcal{L}(\theta) = \frac{1}{|\Omega_T|}\sum_{i \in \Omega_T} \frac{1}{N_i}\|m_i(y_i - f_\theta(x_i))\|_1,
\end{equation}
where $x_i$ is the stack of one or more adjacent input MR image slices, $y_i$ is the corresponding slice from the ground truth CT, $m_i$ is the corresponding mask (see Fig.\ \ref{fig:inputs_examples}) and $N_i$ is the number of pixels in the $i$-th image. The loss function was evaluated stochastically for each optimisation step, by drawing a random mini-batch of size 32 from the training set. The \textit{Adam} optimiser \cite{kingma2014adam} was used along with reduce on plateau scheduling with $\text{patience} = 5$, $\text{factor} = 0.2$ and a learning rate reducing from $10^{-4}$ to $10^{-6}$. The networks were trained for 300 epochs on an NVIDIA Tesla P40 GPU, with 24 GB RAM.

\subsection{\label{sec:classical_zte_mapping}Classical ZTE mapping}

For comparison with the learned pCT mappings, a direct conversion of the ZTE images to pCT was also implemented following \cite{wiesinger2018zero,wiesinger2016zero}. In this case, bias field correction was applied to the ZTE images using the N4ITK method within 3D Slicer (V4.11.20210226). Voxel intensities for each image were individually normalised based on the soft-tissue peak in the image histogram to give a soft-tissue intensity of 1. A skull mask was generated by thresholding the voxel intensities between 0.2 and 0.75 (bone/air and bone/soft-tissue), taking the largest connected component, and then filling the mask using morphological operations. Within the skull mask, ZTE values were mapped directly to CT HU using the linear relationship $\text{CT} = -2085 \text{ ZTE} + 2329$. This relationship was calculated by taking the first principal component of the density plot of ZTE values vs CT values within the skull. Outside the skull mask, air and soft-tissue were assigned values of -1000 and 42 HU, respectively.

\subsection{Evaluation of the pseudo-CT images}

The three different methods for generating pCT images were evaluated by comparing the generated 3D pCT volumes for the 14 subjects in the test set against the corresponding ground truth CTs. Image intensities were compared using mean absolute error (MAE) and the root mean squared error (RMSE). Both metrics were evaluated across the whole head (comparing voxels within the subject-specific head mask) and in the skull only (using a skull mask derived by thresholding the ground truth CTs and combining with the head mask to exclude bones outside the neurocranium). For convenience, the different generated pCTs are referred to as tCT (for the learned pCT mapped from a T1w image), zCT (for the learned pCT mapped from a ZTE image), and cCT (for the pCT directly converted from a ZTE image using classical image processing techniques as explained in Sec.\ \ref{sec:classical_zte_mapping}).

\subsection{Ultrasound simulations using k-Wave}
\label{subsec:ultrasound_simulations}

Acoustic simulations were performed using the open-source k-Wave toolbox \cite{treeby2010k,treeby2012modeling}. CT and pCT image pairs were converted to medium property maps as follows. First, the images were resampled to the simulation resolution (0.5 mm) using linear interpolation and then cropped. The images were then segmented into skull, skin, brain, and background regions using intensity-based thresholding along with morphological operations. Within the soft-tissue, reference values were assigned for the acoustic properties. Within the skull, the sound speed and density were mapped directly from the image values in Hounsfield units. The density was calculated using the conversion curve from \cite{schneider1996calibration} (using the \verb=hounsfield2density= function in k-Wave). The sound speed $c$ within the skull was then calculated from the density values $\rho$ using a linear relationship of $c = 1.33 \rho + 167$ \cite{marquet2009non}. 
A constant value of attenuation within the skull was used. \review{This general approach to mapping the acoustic properties from CT has been widely used in the literature, and generally compares well with experimental and clinical measurements} 
\cite{mcdannold2020predicting, deffieux2010numerical, chauvet2013targeting}.
Note, other mappings from CT images to acoustic properties are also possible (e.g. \cite{pichardo2010multi}). However, as the same mapping is used for all image sets, this choice does not strongly influence the simulated results.

The simulations were based on the NeuroFUS CTX-500 4-element annular array transducer (Sonic Concepts, Bothell, WA). The transducer was modelled using a staircase-free formulation \cite{wise2019representing} using nominal values for the radius of curvature (63.2 mm) and element aperture diameters (32.8, 46, 55.9, 64 mm). Simulations were run at 6 points per wavelength (PPW) in water and 60 points per period (PPP), which was sufficient to reproduce the relevant benchmark results (\verb=PH1-BM7-SC1=) reported in \cite{aubry2022benchmark} with less than 0.2 \% difference in the maximum pressure and no difference in the focal position. The transducer was driven using a continuous sinusoidal driving signal at 500 kHz until steady state was reached.

Simulations were performed for the 14 skulls that formed the test set. For each skull, 4 transducer positions were used targeting the occipital pole of the primary visual cortex and the hand knob of the primary motor cortex in both hemispheres (giving a total of 56 comparisons for each pCT image type). 
\review{These positions are two common targets for TUS studies. The variation in the skull bone thickness for the V1 targets is typically greater due to the internal and external occipital protuberance.}
The target positions were first identified on the \verb=MNI152_T1_1mm= magnetic resonance imaging template brain \cite{fonov2009unbiased,fonov2011unbiased}. Approximate positions of the targets on the individual skulls were then calculated by registering the MR images with the MNI template and using the transformation matrix to map the target positions back to the individual skulls. For comparison between the CT and pCT image pairs, all other simulation parameters were kept identical except the image used to derive the acoustic property maps.

The calculated acoustic pressure fields for each CT/pCT pair were compared using the focal metrics outlined in \cite{aubry2022benchmark} using the code available from \cite{intercomparisoncode}. Briefly, the magnitude and position of the spatial peak pressure within the brain were compared, along with the -6 dB focal volume.

\section{Results}

\subsection{\label{sec:results_pct}Pseudo-CT Images}

Results for the MAE and RMSE for the generated 3D pCT images against the ground truth CT images are given in Table \ref{tab:mae_rmse}. The zCT images generated using the learned mapping from ZTE have significantly lower error values across both metrics and both regions (complete head and skull only) compared to the tCT and cCT images. This is not surprising given (1) the ZTE image visually contains significantly more information about the morphology of the skull bone compared to the T1w image, and (2) the network has significantly more power to learn an adaptive mapping to pCT compared to the fixed mapping used for the cCT. In the skull, the cCT images slightly out-perform the tCT images. This is possibly because the additional bone information available to the cCT (but not the tCT) outweighs the additional predictive power available to the tCT (but not the cCT). Overall, the error values are similar to those discussed in Sec.\ \ref{sec:introduction}.

\begin{table}[]
\centering
\caption{Mean absolute error (MAE) and root mean squared error (RMSE) for the 3D pCT images generated using the 14 subjects in the test set against the ground truth CT images. }
\label{tab:mae_rmse}
\begin{tabular}{cc|cc}
\toprule
mask & & \textbf{MAE} (HU) & \textbf{RMSE}  (HU) \\
\midrule
& tCT & 133 $\pm$ 46 & 288 $\pm$ 83 \\
& zCT & \textbf{83} $\pm$ 26 & \textbf{186} $\pm$ 55  \\
\multirow{-3}{*}{\textbf{head}}  & cCT & 145 $\pm$ 35 &     350  $\pm$  62\\ 
\midrule
& tCT & 398 $\pm$ 116 & 528 $\pm$ 144 \\
& zCT & \textbf{222} $\pm$ 78 & \textbf{305} $\pm$ 102\\
\multirow{-3}{*}{\textbf{skull}} & cCT & 336 $\pm$ 80 & 465 $\pm$ 101 \\
\bottomrule
\end{tabular}
\end{table}

\begin{figure}[h!]
    \centering
     \includegraphics[width=0.48\textwidth]{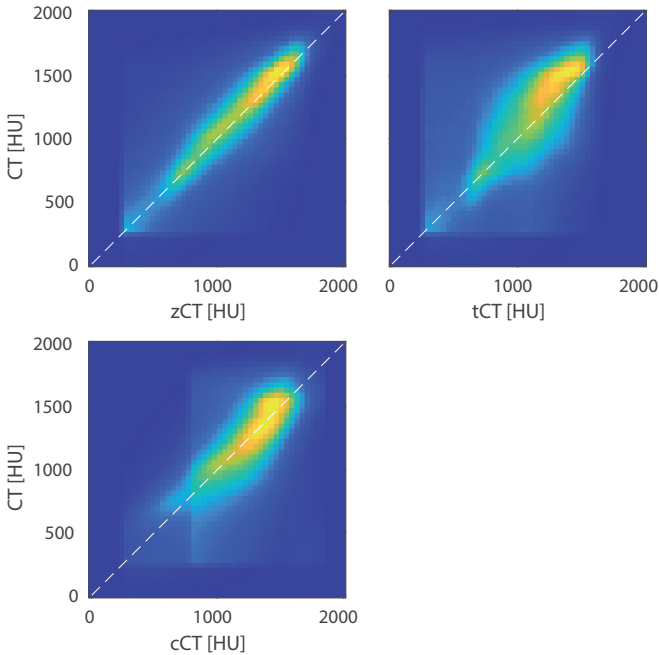}
    \caption{Density plot showing the correlation between the pCT and ground truth CT images for the test set. The white lines show $y = x$.}
    \label{fig:density_plots}
\end{figure}

Figure \ref{fig:density_plots} shows the correlation between the CT and pCT values within the skull mask (as identified from the ground truth CT). There is a good correlation observed for all pCT images, with the lowest spread for the zCT matching the results given in Table \ref{tab:mae_rmse}. The good fit observed for the cCT values demonstrates that a linear mapping from ZTE is a reasonable choice. The robustness of these correlations to variations in the specific MR sequence parameters (along with processing parameters such as the debiasing settings) still needs to be explored further.

\begin{figure*}[t!]
    \centering
    \includegraphics[width=\textwidth]{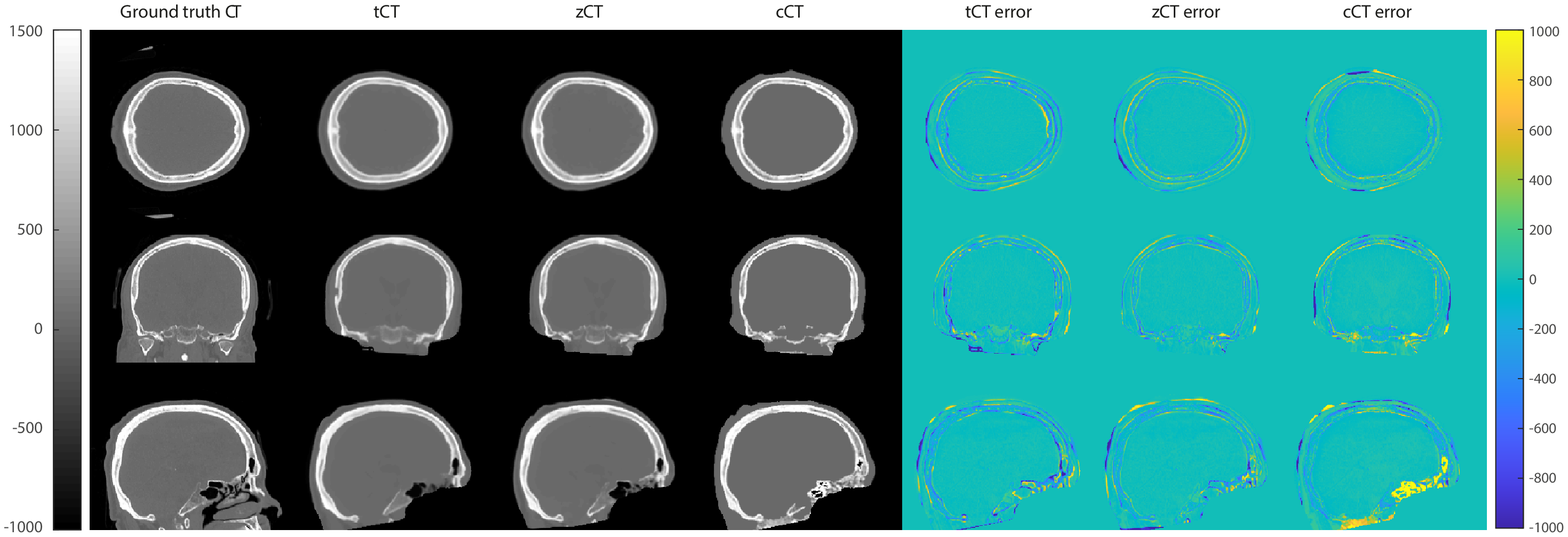}
    \caption{Example of pCT and error maps for a single subject in the test set. The pseudo-CTs are registered to the ground truth CT before comparison.}
    \label{fig:output_pct_example}
\end{figure*}

\begin{figure*}[h!]
    \centering
    \includegraphics[width=\textwidth]{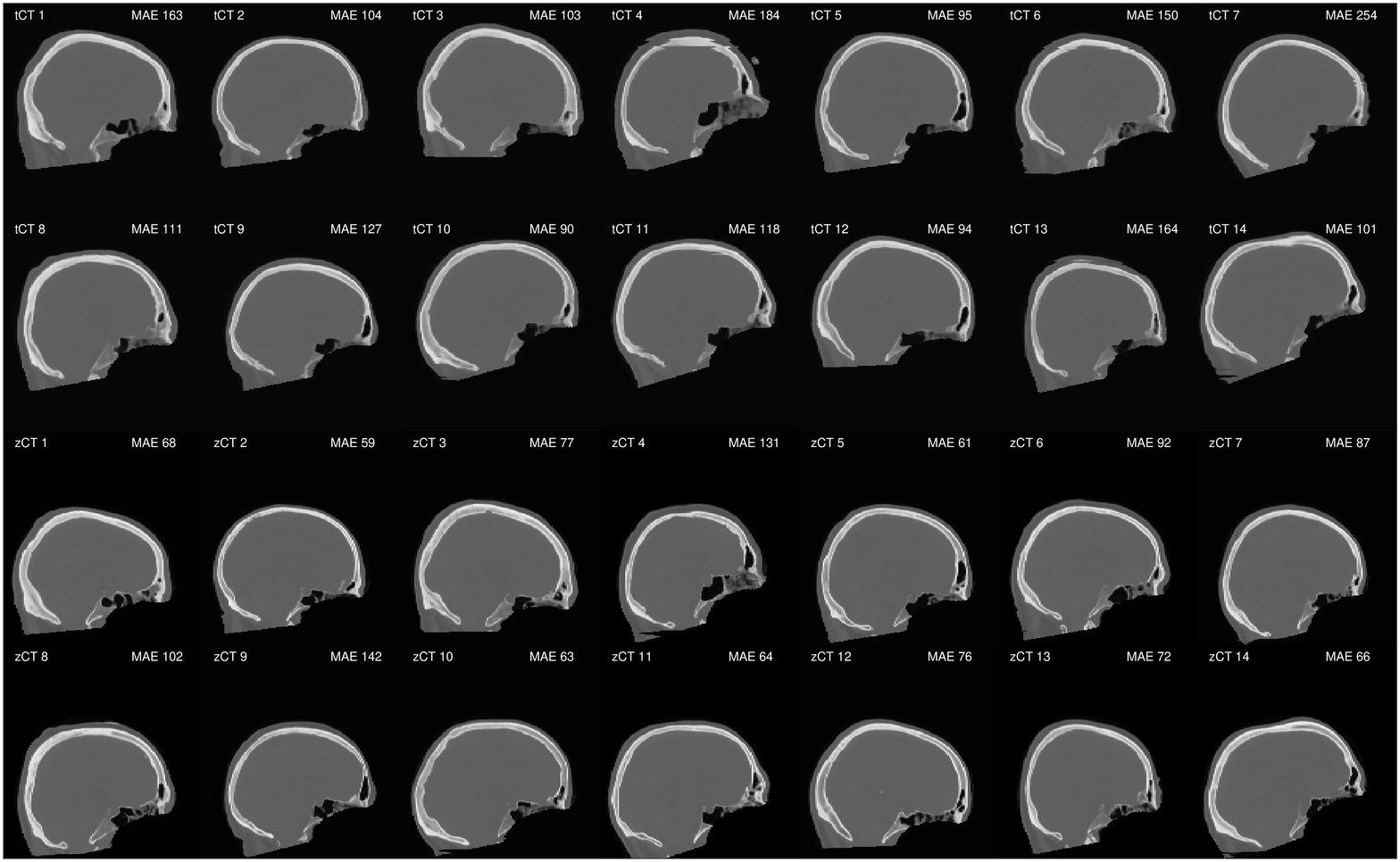}
    \caption{Middle sagittal slice for the learned pseudo-CT images for the 14 test subjects. The mean absolute error (MAE) value reported is calculated for the volume inside the head mask. A small number of the learned images mapped from T1w MR images have discontinuous skull boundaries.}
    \label{fig:all_pcts}
\end{figure*}

Examples of the generated pCT images and corresponding error maps against CT for one subject from the test set are shown in Fig.\ \ref{fig:output_pct_example}. The central sagittal slice through all 14 subjects along with individual MAE values are shown in Fig.\ \ref{fig:all_pcts}. Overall, there is a good quantitative agreement between the pCT and ground truth CT images. The biggest differences occur at the brain-skull and skull-skin boundaries, consistent with previous studies \cite{han2017mr}. This is primarily due to the imperfect registration between the MR and CT images. There are two contributing factors. First, the skin-air interfaces are usually in slightly different places physically, e.g. due to the mobility of the skin and differences in the subject positioning during the MR and CT scans. Second, there are differences in the rigid brain-skull boundaries due to geometric distortions in the images that are not corrected by the affine transformation used in the registration step. This is evident in the difference plots for the cCT images, which uses a simple thresholding of the ZTE image to obtain the skull boundaries. The same misalignments are also apparent in the training/validation images. Improving the registration step, for example by learning the registration as part of training the network \cite{kong2021breaking} or by iterating between training and re-registration of the generated pCT images, will likely improve the predictive capabilities of the learned mappings, and will be the subject of future work.

As discussed in Sec.\ \ref{sec:network_arch}, the input to the network was 11 consecutive 2D slices in the transverse direction. For the tCT images, this significantly reduced the occurrence of discontinuities in the skull between slices in the out-of-plane (e.g. sagittal) direction. The discontinuities occur because of the difficulty in consistently identifying the outer skull boundary in the 2D T1w image slices, which is improved by providing the network with local 3D information.
However, even with 11 slices, visible discontinuities were still observed for a small number of subjects in the test set (tCT 4, 6, and 13 in Fig.\ \ref{fig:all_pcts}). The equivalent network trained with a single input slice had similar MAE and RMSE values to those given in Table \ref{tab:mae_rmse}, while the occurrence of local discontinuities in the skull was much worse. Thus, this feature does not seem to be captured by the error metrics used during training. This motivates using 3D networks in the future, following related work \cite{fu20192d3d} which has shown that training on the full 3D skull volumes can produce better results, albeit at a higher computational cost. The same discontinuities were not observed for the zCT images, even when using a single slice as input to the network.

Although not formally investigated in this study, it is interesting to note that many of the subjects had small calcifications within the deep structures of the brain visible on the CT images (e.g., due to calcification of the choroid plexus). Some, but not all, of these were visible on the zCT images in the test set, and none were visible on the tCT images. Further work is needed to quantify the ability of the learned mappings to correctly reconstruct calcifications.




\subsection{Acoustic Simulations}

\begin{figure}[h!]
    \centering
    \includegraphics[width=0.5\textwidth]{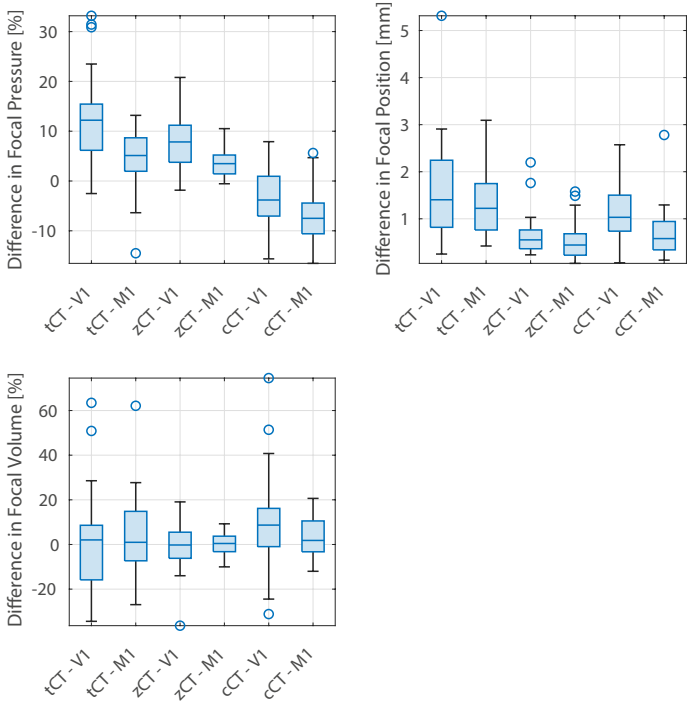}
    \caption{Differences in the focal pressure, focal position, and focal volume for acoustic simulations using different pseudo-CT images against simulations using a ground truth CT. Results are divided into two sets, targeting the visual cortex (V1) and motor cortex (M1).}
    \label{fig:acoustic_metrics}
\end{figure}

Results for the acoustic simulations are summarised in Table \ref{tab:acoustic_metrics} and Fig. \ref{fig:acoustic_metrics}. 
Subjects 4 and 6 from the tCT dataset were excluded from the simulation evaluation as they displayed discontinuities on the top of the skull which made them unsuitable for acoustic simulations. Similarly, cCT from subject 10 was excluded as the predicted skull was not continuous due the thresholding value not being suitable for this subject.
Across all pCT images and target locations, the mean differences in the simulated focal pressure, focal position, and focal volume were 7.3 $\pm$ 5.7 \%, 0.99  $\pm$ 0.79 mm, and 11 $\pm$ 12 \% (to two significant figures). The best results were obtained when using the zCT images for motor targets, where the equivalent differences were 3.7 \%, 0.5 mm, and 3.9 \%. These values compare well with the differences observed in experimental repeatability  \cite{martin2019investigation} and numerical intercomparison \cite{aubry2022benchmark} studies.  For reference, the focal volume simulated in water is 3.9 mm wide and 24 mm long.

Between the different pCT images, the simulations based on the zCT images generally had the lowest average errors, as well as the smallest variation (see the interquartile range shown using the blue boxes in Fig.\ \ref{fig:acoustic_metrics}). The simulations based on both the zCT and cCT images consistently outperformed the tCT images, demonstrating that skull-specific imaging can improve the accuracy of the predictions. Interestingly, the simulations based on the learned mappings (tCT and zCT) generally over-estimated the focal pressure, while simulations based on the the direct mapping (cCT) generally under-estimated the focal pressure. This may be related to the sharpness of the skull boundaries in the cCT images compared to the learned images resulting in a stronger reflection coefficient.

\begin{figure*}[h!]
    \centering
    \includegraphics[width=0.8\textwidth]{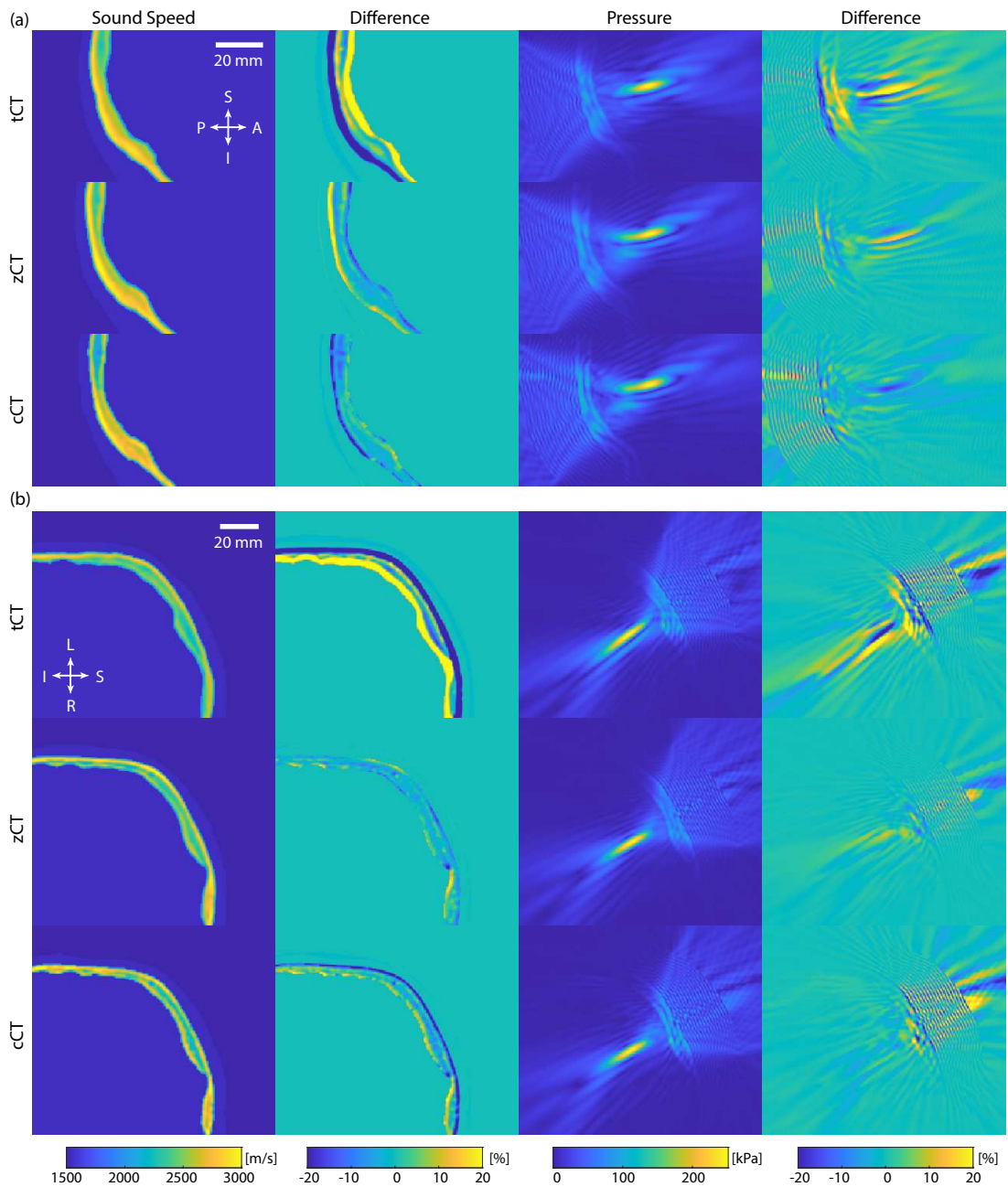}
    \caption{Sound speed maps and simulated acoustic field for a subject from the test set for targets in the (a) visual cortex (V1) and (b) motor cortex (M1). The difference plots show the difference against the ground truth generated using the real CT images. The V1 and M1 results show sagittal and coronal slices through the spatial peak pressure, respectively.}
    \label{fig:sim_VC}
\end{figure*}

Between the two targets, the errors for the motor cortex were lower than the visual cortex, except for the focal pressure metric for the cCT images. This is expected, as the shape of the skull is generally more variable close to the visual cortex due to the internal and external occipital protuberance, which can result in stronger aberrations to the acoustic field. Examples from one test subject of the generated sound speed maps and acoustic field distributions are given in Fig.\ \ref{fig:sim_VC}.

\begin{table}[t]
    \centering
    \caption{acoustic metrics rounded to 1 decimal place}
    \label{tab:acoustic_metrics}
    
    \begin{tabular}{c|c|c|c|c|c|c|c}
    \toprule
    \multirow{2}{*}{\textbf{Target}} & \multirow{2}{*}{\textbf{CT}} &  \multicolumn{2}{c|}{\textbf{focal pressure}} & \multicolumn{2}{c|}{\textbf{focal position}} & \multicolumn{2}{c}{\textbf{focal volume}}\\
    \multirow{2}{*}{} & \multirow{2}{*}{} &  \multicolumn{2}{c|}{\textbf{(\%)}} & \multicolumn{2}{c|}{\textbf{(mm)}} & \multicolumn{2}{c}{\textbf{(\%)}}\\
    \textbf{} & & \textbf{mean} & \textbf{std} & \textbf{mean} & \textbf{std} & \textbf{mean} & \textbf{std} \\
    \midrule
    \multirow{3}{*}{V1} &
    tCT & 13.1 & 9.0 & 1.6 & 1.1 & 17.2 & 15.8\\
    & zCT & 7.6 & 4.9 &  \textbf{0.7} & 0.4 & \textbf{7.4} & 7.1\\
    & cCT  & \textbf{5.7} & 4.0 &  1.1 & 0.6 & 16.8 & 17.6\\
    \midrule
    \multirow{3}{*}{M1} &
    tCT  &  6.7 & 3.5 & 1.4 & 0.8 &  13.0 & 13.3\\
    & zCT &  \textbf{3.7} & 2.7 & \textbf{0.5} & 0.4 & \textbf{3.9} & 2.8 \\
    & cCT &  7.8 & 3.8 & 0.7 & 0.5 & 7.4 & 5.7 \\
    \midrule
    \multirow{3}{*}{All} &
    tCT & 9.9 & 7.5 &  1.5 & 1.0 & 15.1 & 14.6\\
    & zCT & \textbf{5.7}  & 4.4 & \textbf{0.6} & 0.4 & \textbf{5.7} & 5.6\\
    & cCT & 6.7 & 4.0 & 0.9 & 0.6 & 12.1 & 13.8\\
    \bottomrule
    \end{tabular}
\end{table}

\section{Summary and Discussion}

Three different approaches for generating pCT images from MR images were investigated in the context of treatment planning simulations for transcranial ultrasound stimulation. A convolutional neural network (U-Net) was trained to generate pCT images using paired MR-CT data with either T1w or ZTE MR images as input. A direct mapping from ZTE to pCT was also implemented based on \cite{wiesinger2018zero}. For the image-based metrics, the learned mapping from ZTE gave the lowest errors, with MAE values of 83 and 222 HU in the whole head and skull-only, respectively. The significant improvement compared to the mapping from T1w images demonstrates the advantage of using skull-specific MR imaging sequences.

\review{Acoustic simulations were performed using the generated pCT images for an annular array transducer geometry operating at 500 kHz. The transducer was targeting either the left or right motor cortices, or the left or right visual cortices. The choice of transducer geometry and brain targets was motivated by devices and targets used in the rapidly-growing literature on human transcranial ultrasound stimulation (TUS).}

\review{Simulations using the pCT images based on ZTE showed close agreement with ground truth simulations based on CT, with mean errors in the focal pressure and focal volume less than 7 \% and 9 \% respectively, and mean errors in the focal position less than 0.8 mm (see Table} \ref{tab:acoustic_metrics}).
\review{Errors in simulations using the learned pCT images mapped from T1w images were higher, but still may be acceptable depending on the accuracy required. These results demonstrate that acoustic simulations based on mapping pCT images from MR can give comparable results to simulations based on ground truth CT.} 

\review{For context, in the motor cortex, the smallest relevant target area for stimulation could reasonably be considered the cortical representation of a single hand muscle.  A recent transcranial magnetic stimulation (TMS) study suggests that the functional area of the first dorsal interosseous (FDI), a muscle commonly targeted in motor studies, is approximately 26 mm$^2$ }\cite{reijonen2020spatial}.\review{In the primary visual cortex, the mean distance between approximate centres of TMS-distinct visual regions (V1 and V2d) is 11 mm}\cite{salminen2012selective},\review{suggesting a similar necessary resolution to M1. The errors in the focal position using the pCT images are significantly less, which should allow for accurate targeting of the intended structures.}

\review{Interestingly, there was not a strong correlation between the pixel-based image errors (MAE and RMSE) and the errors in the predicted acoustic field. This highlights the importance of running acoustic simulations when developing image-to-image translation methods for TUS, and potentially calls for the use of acoustic-based loss functions.}\review{The use of MR images for treatment planning is particularly important for neuroscience studies in healthy populations, where obtaining ethical approval to acquire CT images is usually problematic. Even in a clinical setting, there may be benefits to using MR-based acquisitions when the therapy is performed under MR-guidance (for example, multi-modal image registration errors are reduced}\cite{edmund2017review}).\review{These results may also be of interest in other areas where pseudo-CTs are used, for example, PET-MR and radiotherapy planning.}

\review{Note, in a clinical setting, the regulatory approval of learned mappings remains an important open question. However, we demonstrate that the learned pseudo-CTs from T1w and ZTE can be used for acoustic simulations which have comparable accuracy to those based on CTs.}





\section*{Acknowledgments}
This work was supported by the Engineering and Physical Sciences Research Council (EPSRC), UK, grant number EP/S026371/1 and the UKRI CDT in AI-enabled Healthcare Systems, grant number EP/S021612/1.

\bibliography{bibliography}{}
\bibliographystyle{IEEEtran}


\end{document}